\documentclass[aps,prd,twocolumn,superscriptaddress,preprintnumbers]{revtex4-2}
\usepackage[normalem]{ulem}
\usepackage{amsmath}
\usepackage{amsfonts}
\usepackage{amssymb}
\usepackage{graphicx}
\usepackage[colorlinks=True,citecolor=myblue,linkcolor=myorange,urlcolor=mygreen]{hyperref}
\usepackage{hypcap}
\usepackage{braket}
\usepackage{bm}
\usepackage{bbm}
\usepackage{tikz}
\usepackage[export]{adjustbox}
\usepackage{xcolor}
\pagecolor{white}
\usepackage{mathtools}
\mathtoolsset{centercolon}
\usepackage{mathrsfs}

\definecolor{myblue}{RGB}{0,0,130}
\definecolor{myorange}{RGB}{130,50,0}
\definecolor{mygreen}{RGB}{0,130,0}

\usepackage{tikz}
\usetikzlibrary{arrows,decorations.pathmorphing}

\begin{document}

\title{Quantum Algorithm to Prepare Quasi-Stationary States}
\author{Samuel J. Garratt}
\affiliation{Department of Physics, University of California, Berkeley, CA 94720, USA}
\author{Soonwon Choi}
\affiliation{Center for Theoretical Physics, Massachusetts Institute of Technology, Cambridge, MA 02139, USA}

\preprint{MIT-CTP/5734}

\begin{abstract}
Quantum dynamics can be analyzed via the structure of energy eigenstates. However, in the many-body setting, preparing eigenstates associated with finite temperatures requires time scaling exponentially with system size. In this work we present an efficient quantum search algorithm which produces \emph{quasi-stationary states}, having energies supported within narrow windows of a dense many-body spectrum.
In time scaling polynomially with system size, the algorithm produces states with inverse polynomial energy width, which can in turn be used to analyze many-body dynamics out to polynomial times. 
The algorithm is based on quantum singular value transformations and quantum signal processing, and provides a quadratic speedup over measurement-based approaches.
We discuss how this algorithm can be used as a primitive to investigate the mechanisms underlying thermalization and hydrodynamics in many-body quantum systems. 
\end{abstract}

\maketitle

One of the natural applications of quantum technology is to probe basic science questions for which no other means provides scalable solutions.
An example includes understanding \emph{quantum thermalization}: how an isolated quantum system, evolving under reversible unitary dynamics, approaches a high entropy thermal state \cite{deutsch1991quantum,srednicki1994chaos,rigol2008thermalization,dalessio2016quantum,
kaufman2016quantum,choi2016exploring,schreiber2016observation}. The apparent contradiction between the reversibility of unitary dynamics and the second law of thermodynamics is resolved by the fact that the evolving quantum states become highly entangled \cite{calabrese2005evolution,kim2013ballistic,nahum2017quantum}. Entanglement growth, however, also pushes the exact description of quantum thermalization beyond the reach of classical simulations~\cite{vidal2003efficient}, which makes it challenging to study this phenomenon.

This situation raises the question of how quantum computers can be used to unravel the mechanisms underlying quantum thermalization. It is widely expected that this process can be generically understood through the lens of the eigenstate thermalization hypothesis (ETH) ~\cite{deutsch1991quantum,srednicki1994chaos,deutsch2018eigenstate}, whose strong form asserts that many-body eigenstates resemble random vectors and mimic thermal states locally. The implication is that thermalization is a consequence of ``dephasing'': weakly-entangled initial states correspond to highly structured superpositions of eigenstates \cite{foini2024out} and, as phase differences develop between eigenstates with different energies, the structure in this superposition is lost. A thermalized state then corresponds to a random superposition of eigenstates~\cite{mark2024maximum}.

The (strong) ETH is appealing in its simplicity but it is not implied by the observation of thermalization from any single initial state \cite{biroli2010effect,palma2015necessity,mori2017thermalization,shiraishi2018analytic,harrow2023thermalization}.
Unfortunately, testing the ETH directly is a difficult if not impossible task: constructing and probing a many-body eigenstate at finite energy density typically requires exponential time even with a quantum computer~\cite{atia2017fast,abrams1999quantum,poulin2009preparing,lin2020optimal}.

If our aim is to analyze thermalization it should not be necessary to invoke exponentially difficult quantum computation. In physical settings this process is typically controlled by hydrodynamic modes, which relax over polynomial time scales \cite{chaikin1995principles}. The implication is that thermalization arises from the dephasing of eigenstates with inverse polynomially, rather than inverse exponentially, separated energies. 
This observation motivates the decomposition of an initial state into a superposition of a set of ``effective eigenstates'' with inverse polynomial energy width --- we dub these \emph{quasi-stationary states} (QSSs).

\begin{figure}
\includegraphics[width=0.47\textwidth]{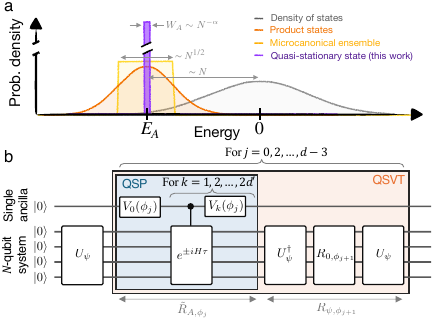}
\vspace{-5pt}
\caption{
(a) Our algorithm prepares QSSs --- quantum states supported on a narrow energy window centered at given $E_A$ and of width $W_A$ for a many-body system of size $N$.
(b) The circuit diagram of our algorithm, which involves only $N+1$ qubits and simple iterated operations.}
\label{fig:dos}
\end{figure}

In this work, we show how state-of-the-art techniques such as quantum signal processing (QSP)~\cite{low2017optimal} and quantum singular value transformations (QSVT)~\cite{gilyen2019quantum,martyn2021grand} can be synthesized with a system's natural dynamics in order to create QSSs (see Fig.~\ref{fig:dos}). Our algorithm is based on quantum search \cite{grover1996fast,brassard2002quantum} and benefits from a polynomial speedup over previous approaches to accomplish related tasks~\cite{yang2020quantum,vasilyev2020monitoring,irmejs2023efficient}. We leverage QSP and QSVT to substantially reduce resource requirements and provide rigorous bounds on accuracy. 

The technical ingredients necessary to realize our algorithm are (i) single-qubit-controlled Hamiltonian evolution; (ii) a multiqubit Toffoli gate~\cite{barenco1995elementary}; and (iii) arbitrary single-qubit rotations. 
Several multiqubit couplings have been proposed or already demonstrated based on interactions native to Rydberg atoms \cite{levine2019parallel,evered2023high,young2021asymmetric}, trapped ions \cite{monz2009realization,katz2023demonstration,espinoza2021high}, and systems of superconducting qubits \cite{kim2022high,khazali2020fast}.
Therefore, our algorithm is a promising candidate to produce practical benefits from near-term devices.

\emph{Quasi-stationary states} --- We begin by introducing the setting, our notation, and the definition of QSSs. 
We consider a system of $N$ qubits with Hamiltonian $H = \sum_{a=0}^{D-1} E_a \ket{E_a}\bra{E_a}$, where $E_a$ and $\ket{E_a}$ are energy eigenvalues and eigenvectors in the Hilbert space $\mathcal{H}$ of dimension $D=2^N$.
An energy window $\mathcal{H}_A$, specified by center $E_A$ and width $W_A$, is the subspace spanned by eigenstates with $|E_a-E_A|<W_A/2$. The complement of $\mathcal{H}_A$ is denoted $\mathcal{H}_{\bar{A}}$ such that $\mathcal{H} = \mathcal{H}_A \oplus \mathcal{H}_{\bar{A}}$, and the projectors onto these spaces are $\Pi_A$ and $\Pi_{\bar{A}}$. We are primarily interested in $E_A$ located at finite energy density (but not necessarily at infinite temperature) and $W_A$ inverse polynomially small in $N$, which is assumed below. A state $\ket{\psi_A}$ is a QSS in $\mathcal{H}_A$ if $|\Pi_A \ket{\psi_A}|^2 = 1-|\Pi_{\bar{A}} \ket{\psi_A}|^2 = 1-\epsilon$ where the population error $\epsilon \ll 1$. In this work we focus on $\epsilon$ that is at most inverse polynomially small in $N$. 

QSSs with inverse polynomial widths $W_A$ provide a way to analyze dynamics out to polynomial times. In particular, the dynamics of an observable $\mathcal{O}$ following a quench from $\ket{\psi}$ can be approximated by $\braket{\mathcal{O}(t)} \simeq \sum_{AA'}\sqrt{p_A p_{A'}}\braket{\psi_A|\mathcal{O}|\psi_{A'}}e^{i(E_A-E_{A'})t}$. The normalized QSSs in this expression are $\ket{\psi_A}=\Pi_A\ket{\psi}/\sqrt{p_A}$. These are specific kinds of energy-filtered states \cite{banuls2020entanglement,yang2020quantum,lu2021algorithms,morettini2024energy}, here supported in nonoverlapping windows, i.e. $|E_{A}-E_{A'}| \geq (W_A+W_{A'})/2$. The error in this `QSS decomposition' of $\braket{\mathcal{O}(t)}$ is $O(Wt)$ for $Wt \ll 1$, where $W = \text{max}_A W_A$ and $\sum_A p_A \simeq 1$ \cite{SM}.

\emph{Setup} --- Our aims are to prepare such QSSs for a given $A$ and generic local $H$, as well as to determine the quantities $\braket{\psi_A|\mathcal{O}|\psi_{A'}}$ for an observable $\mathcal{O}$. The access to $H$ is given as our ability to realize the controlled evolution
\begin{align}
\textrm{C}U_{H}(\tau) = \ket{0}\bra{0} \otimes e^{-i H \tau}+ \ket{1}\bra{1} \otimes  \mathbbm{1}
\end{align}
for a sufficiently small $\tau$ satisfying $|E_{D-1} - E_0| \tau \leq 2\pi$, as well as the inverse of this operator.
This unitary acts on one ancilla qubit as control and $N$ system qubits as target. 
Using $\textrm{C}U_H$
and unitary operations acting only on the control qubit, we will show how to implement nonlinear transformations of $U_H\equiv e^{-iH\tau}$, which will be parts of our algorithm.

We additionally assume access to unitaries $U_{\psi}$ and $U_{\psi}^{-1}$ (of depth at most polynomial in $N$) such that $\ket{\psi}= U_{\psi} \ket{0}^{\otimes N}$ with $p_A = |\Pi_A\ket{\psi}|^2$ inverse polynomial in $N$. For the worst-case $H$ and for $\ket{\psi}$ with finite energy density $E_{\psi}/N=\braket{H}/N$, the existence of such $U_{\psi}$ would be at odds with the Quantum PCP conjecture~\cite{aharonov2013guest} and the assumption that the complexity classes QMA and QCMA are distinct~\cite{aharonov2002quantum,aaronson2007quantum}. However, for geometrically local $H$, we can create $\ket{\psi}$ with any finite energy density using finite-depth $U_{\psi}$  \cite{hastings2012trivial}. Moreover, for physical Hamiltonians it is widely expected that such short-range correlated $\ket{\psi}$ has nonzero (but exponentially small) overlap with almost all eigenstates with energies $E_a$ such that $|E_a-E_{\psi}| \lesssim W_{\psi}$, where $W_{\psi}$ is the energy width of $\ket{\psi}$ with respect to $H$. The implication is that $p_A \sim W_A/W_{\psi}$. Assuming access only to $\textrm{C}U_{H}$ and $U_{\psi}$ we will show how to efficiently create a QSS $\ket{\tilde \psi_A}$ that approximates $\ket{\psi_A}=\Pi_A\ket{\psi}/\sqrt{p_A}$.

\emph{Overview} --- The key idea is to view the state preparation as a search in the energy spectrum. Indeed, quantum search can be generally viewed as unitary transformations from a simple input state to an output state belonging to a target space.
To this end, let us first outline our strategy in terms of Grover's algorithm~\cite{grover1996fast,brassard2002quantum}. The algorithm requires as an ingredient a unitary operation which `marks' components of the state within the target space, here $\mathcal{H}_A$; this is effected by a reflection $R_A=\mathbbm{1}-2\Pi_A$. One additionally requires the reflection $R_{\psi}=\mathbbm{1}-2\Pi_{\psi}$ about the initial state, where $\Pi_{\psi}=\ket{\psi}\bra{\psi}$. 

Grover's algorithm operates on the geometric principle that the composition $R_{\psi}R_A$ of the two reflections defines a rotation in a two-dimensional space spanned by $\ket{\psi_A}$ and $\ket{\psi_{\bar{A}}}=\Pi_{\bar{A}}\ket{\psi}/\sqrt{1-p_A}$. The rotations are in increments of $2\chi$, where $\sin \chi = \sqrt{p_A}$, and since $\braket{\psi_{\bar{A}}|\psi}=\cos\chi$ one requires $M=\lfloor \pi/(4\chi)\rfloor$ rotations in order to maximize the amplitude in the target space $\mathcal{H}_A$. For $p_A \ll 1$ we then have $M \sim 1/\sqrt{p_A}$. This corresponds to a quadratic speedup relative to a ``classical'' preparation of $\ket{\psi_A}$ based on measuring $\Pi_A$ \cite{yang2020quantum,vasilyev2020monitoring}, which typically requires $\sim 1/p_A$ attempts.

Our approach is similar to Grover's algorithm at a conceptual level, but it addresses two central challenges.
First, Grover search requires knowing $M$ and hence $p_A$ in advance, and moreover suffers an error whenever $\pi/(4\chi)$ is not an integer. In our setting, $p_A$ is not known precisely. Second, one needs to efficiently implement operations that `mark'  $\mathcal{H}_A$ and $\ket{\psi}$ such as $R_A$ and $R_\psi$.

The first challenge can be overcome using a generalization of Grover search known as fixed-point (FP) quantum search \cite{yoder2014fixed}. 
FP search runs in deterministic time, succeeds with high probability, and achieves high fidelity independent of $p_A$ as long as $p_* < p_A$ for a predetermined threshold $p_*$. Both FP search and Grover search can be viewed as QSVTs, as we discuss below. Under the QSVT framework, our second challenge translates to efficient implementations of `generalized' reflections
\begin{align}
R_{\mathsf{x},\phi} &\equiv e^{-i\phi} (\mathbbm{1} - \Pi_\mathsf{x}) + e^{i\phi} \Pi_\mathsf{x},
\end{align}
where $\mathsf{x}=\psi$ or $A$. The reflections used in Grover search correspond to setting constant $\phi=\pi/2$ (up to a global phase), while FP search uses different values of $\phi$ in different iterations.

Implementing $R_{\psi,\phi}$ is straightforward given access to $U_\psi$, since $R_{\psi,\phi} = U_\psi  R_{0,\phi} U_\psi^\dagger$ where $R_{0,\phi} = e^{-i\phi} (1-\Pi_0) + e^{i\phi} \Pi_0$ and $\Pi_0$ is the projector onto state $\ket{0}^{\otimes N}$. The operation $R_{0,\phi}$ can then be constructed from a Toffoli gate with $N$ control qubits.
To implement $R_{A,\phi}$ we must utilize $\textrm{C}U_H$. We now show how QSP can be used to construct nonlinear transformations of $e^{-iH\tau}$ which approximate the generalized reflection $R_{A,\phi}$. Following this we outline the FP search based on QSVT \cite{martyn2021grand}.

\emph{QSP} --- 
The gist of generalized QSP, introduced in Ref.~\cite{motlagh2024generalized}, is as follows. When a qubit is subjected $m$ repeated applications of a \emph{signal rotation}, $S_\theta = e^{-i\theta} \ket{0}\bra{0} +  \ket{1}\bra{1}$, interspersed with $m+1$ \emph{control rotations}, $\vec{V} = (V_0, V_1, \dots, V_m)$ with $V_j \in  U(2)$,
the signal $e^{-i\theta}$ can be processed in a nonlinear fashion:
\begin{align}
\label{eqn:qsp}
U_{\vec{V}} \equiv V_m S_\theta \cdots V_2 S_\theta V_1 S_\theta V_0
=
\left[
\begin{array}{cc}
Q(e^{-i \theta}) & \cdot \\
\cdot & \cdot 
\end{array}
\right],
\end{align}
where the processed signal $Q(e^{-i\theta})$ is encoded as the $\ket{0}\bra{0}$ matrix element, and the dots represents the other elements. 
By suitably choosing $\vec{V}$, one can engineer the transformation function $Q(\cdot)$ as an arbitrary complex polynomial of degree at most $m$ subject to the constraint $|Q(e^{-i\theta})|\leq 1$ for real $\theta$ \cite{motlagh2024generalized}. By subsequently acting $n$ times with the inverse signal rotation $S_{-\theta}$ we see that our signal can be instead processed by a Laurent polynomial $Q'(e^{-i\theta}) = e^{in\theta}Q(e^{-i\theta})$ in $e^{-i\theta}$, here a sum of powers of $e^{-i\theta}$ from $e^{in\theta}$ to $e^{-i(m-n)\theta}$. Methods to determine the set of control rotations necessary to implement a desired transformation $Q'(\cdot)$ have been developed in Refs.~\cite{motlagh2024generalized,yamamoto2024robust}.

Importantly the signal $\theta$ need not be classically valued: Here it is replaced by an operator $H\tau$ acting on $N$ system qubits.
To see this, consider using $\text{C}U_H$ (and its inverse) as the signal rotations in Eq.~\eqref{eqn:qsp} when the system is prepared in an eigenstate $\ket{E_a}$.
Each time $\text{C}U_H$ is applied, the ancilla evolves effectively under $ e^{-iE_a\tau} \ket{0}\bra{0} + \ket{1}\bra{1} $.
This observation, together with linearity, leads to the operator equality:
\begin{align}
\label{eqn:qsp2}
    U_{\vec{V}} = \sum_a \left[
\begin{array}{cc}
Q'(e^{-i E_a \tau }) & \cdot \\
\cdot & \cdot 
\end{array}
\right] \otimes \ket{E_a}\bra{E_a}.
\end{align}
The processed ``matrix element'' is now the operator $ \tilde R_{A,\phi} \equiv (\bra{0}\otimes \mathbbm{1}) U_{\vec{V}} (\ket{0} \otimes \mathbbm{1}) = Q'(e^{-iH\tau})$. This nonlinear tranformation of $e^{-iH\tau}$ will approximate $R_{A,\phi}$.

Concretely, $\tilde R_{A,\phi}$ is realized in three steps: (i) initialize the ancilla in $\ket{0}$, (ii) evolve it under $U_{\vec{V}}$ in Eq.~\eqref{eqn:qsp2}, and (iii) measure the ancilla in the computational basis. If in step (iii) we find $\ket{0}$ we act on the system with $\tilde R_{A,\phi}$.
In the ideal limit, $\tilde{R}_{A,\phi} = R_{A,\phi}$, the measurement always yields $\ket{0}$ since $R_{A,\phi}$ is unitary, and hence (iii) can be omitted.
In practice, the small discrepancy between $R_{A,\phi}$ and $\tilde R_{A,\phi}$ leads to a small probability to measure $\ket{1}$, and our algorithm fails. We bound this probability below.

It remains to find a good polynomial approximation for $R_{A,\phi}$.
Note that $R_{A,\phi}$ can be understood as a transformed Hamiltonian operator: $R_{A,\phi}=r_{A,\phi}(H\tau)$, where the periodic function
\begin{align}
    r_{A,\phi}(\theta) = \left\{ 
    \begin{array}{cc}
    e^{i\phi} & \textrm{ if } |\theta-E_A\tau| \leq W_A\tau, \\
    e^{-i\phi} & \textrm{ if } |\theta-E_A\tau| > W_A\tau.
    \end{array}
    \right. \label{eq:rAphi}
\end{align}
Here $|\theta-E_A\tau|$ should be understood as an angular distance measured around the unit circle, so $r_{A,\phi}(\theta)=r_{A,\phi}(\theta+2\pi)$~\footnote{Note that $r_{A,\phi}$ can be consistently defined on a circle due to the condition $|E_1-E_D|\tau \leq 2\pi$}. We approximate $r_{A,\phi}(\theta)$ with a Laurent polynomial in $e^{-i\theta}$ involving powers from $-d'$ to $d'$, which is a Fourier series $\tilde r_{A,\phi}(\theta) 
= \sum_{\ell=-d}^{d} q_{\phi,\ell} e^{-i \ell \theta}$ such that $\braket{E_a|\tilde R_{A,\phi}|E_a}=\tilde r_{A,\phi}(E_a\tau)$; in the notation below Eq.~\eqref{eqn:qsp2} we have set $m=2d'$, $n=d'$. It is not possible to perfectly resolve the edges of the window $\mathcal{H}_A$ with less than exponential $d'$ because the energy level spacing is exponentially small, and to control the error which arises when $d'$ is polynomial in $N$ we must `blur' the discontinuities over an energy scale $B$, i.e. $\tilde r_{A,\phi}(\theta)$ smoothly changes from $e^{i\phi}$ to $e^{-i\phi}$ over a $\theta$ interval $\sim B\tau$ \cite{SM}. The blurring energy scale $B \ll W_A$, and the polynomial degree $d'$ is constrained by $d'B\tau \gg 1$. Having explained how to approximate $R_{A,\phi}$, we now turn to the FP search algorithm.

\emph{QSVT} --- It is useful to introduce QSVT using Grover search as an example. A QSVT is defined in terms of two projection operators, here $\Pi_A$ and $\Pi_{\psi}$, and the singular value decomposition of their product, here $\Pi_A \Pi_{\psi}=\sin \chi \ket{\psi_A}\bra{\psi}$. 
The singular value $\sin \chi$ can be transformed by inserting an appropriate unitary between the projectors: in the case of Grover search we have $\Pi_A(R_{\psi}R_A)^M\Pi_{\psi}=(-1)^M \sin[(2M+1)\chi]\ket{\psi_A}\bra{\psi}$. Here, the transformed singular value $(-1)^M \sin[(2M+1)\chi]=T_{2M+1}(\sin \chi)$, which is the definition of the degree $2M+1$ Chebyshev polynomial of the first kind $T_{2M+1}(\cdot)$. The FP search is accomplished in a similar manner with a different choice of polynomial.

In general, QSVTs are implemented using unitary operations $U_{\vec{\phi}}$ that are alternating products of $R_{A,\phi}$ and $R_{\psi,\phi}$,
\begin{align}
    U_{\vec{\phi}} = R_{A,\phi_{d-1}} R_{\psi,\phi_{d-2}} \cdots R_{A,\phi_2} R_{\psi,\phi_1} R_{A,\phi_0}, \label{eq:QSVT}
\end{align}
where we identify $\vec{\phi}=(\phi_0,\ldots,\phi_{d-1})$. Remarkably, there exist control parameters $\vec{\phi}$ such that $\Pi_A U_{\phi} \Pi_{\psi}= P(\sqrt{p_A}) \ket{\psi_A}\bra{\psi}$ for any degree $d$ odd polynomial $P(\cdot)$ for which $|P(\cdot)| \leq 1$ when the (real) argument has modulus less than or equal to unity \cite{gilyen2019quantum,martyn2021grand}~\footnote{When the product of projectors has rank greater than unity, all singular values are transformed by the same polynomial}.

The polynomial in FP search is chosen such that any $p_A\geq p_*$ is amplified: $P(\sqrt{p_A})>\sqrt{1-\Delta^2}$ for $p_* \leq p_A \leq 1$, where $\Delta$ is a tunable accuracy parameter~\cite{yoder2014fixed}. The optimal control parameters $\vec{\phi}$ for the FP search are well known (where $\phi_{d-1} = 0$)
and the required polynomial degree $d\geq d_* = 2\lceil\ln(2/\Delta)/(2\sqrt{p_*})\rceil+1$ is controlled by both $p_*$ and the accuracy $\Delta$. 
Hence, the algorithm corresponds to the unitary $U_{\vec{\phi}}=F_{\psi,A}$ with $F_{\psi,A}=\prod_{k=0}^{(d-3)/2} R_{\psi,\phi_{2k+1}}R_{A,\phi_{2k}}$. Altogether, our QSS preparation algorithm is FP search with the operations $R_{A,\phi}$ replaced by their blurred approximations $\tilde R_{A,\phi}$:
\begin{align}
	\tilde F_{\psi,A} = R_{\psi,\phi_{d-2}} \tilde R_{A,\phi_{d-3}} \cdots R_{\psi,\phi_1} \tilde R_{A,\phi_0}. \label{eq:Fsearchapprox}
\end{align}
This results in the QSS $\ket{\tilde \psi_A} \equiv \tilde F_{\psi,A}\ket{\psi}/|\tilde F_{\psi,A}\ket{\psi}|$.
See Fig.~\ref{fig:dos}b.

\emph{Performance} --- 
The performance of our algorithm is assessed along three axes: (i) the population error in the final state, $\epsilon = | \Pi_{\bar A} \ket{\tilde{\psi}_A}|^2$, (ii) the failure probability $p_{\text{fail}}=1-|\tilde F_{\psi,A}\ket{\psi}|^2$, and (iii) the average run-time. 
The target $A$, accuracy $\Delta$, and probability lower bound $p_*$ (which depends on our choice of $U_\psi$), determine the minimum required QSVT degree $d$ \cite{yoder2014fixed}. Our error analysis \cite{SM} then shows how $A$ and $d$ determine the maximum permissible blur $B$, and this constrains the QSP degree $d' \gg [B \tau]^{-1}$, the run-time, and the errors.

First, note that short-range correlated states $\ket{\psi}$ have energy width $W_{\psi}$\,$=$\,$\Theta(N^{1/2})$ with respect to local $H$, so for a window with width $W_A$\,$=$\,$\Theta(N^{-\alpha})$ we assume $p_A$\,$=$\,$\Theta(N^{-1/2-\alpha})$. Setting $\Delta$\,$=$\,$\Theta(N^{-\delta})$ and $p_*$\,$\sim$\,$p_A$ we require $d$\,$=$\,$\Theta(N^{\alpha/2+1/4}\log N)$. The blurring of the generalized reflections $\tilde R_{A,\phi}$ is a key source of error, and the squared amplitude of $\ket{\psi}$ within the energy intervals of width $B$\,$=$\,$\Theta(N^{-\beta})$ centered on $E_A$\,$\pm$\,$W_A/2$ is assumed to be $p_B$\,$=$\,$O(N^{-1/2-\beta+0^+})$, where $0^+$ is a number greater than zero which absorbs logarithmic factors. Allowing for the (possibly coherent) addition of $d$ amplitude errors, each having magnitude $O(N^{-1/4-\beta/2+0^+})$, we find a rigorous upper bound on the population error $\epsilon = O(N^{\alpha-\beta+0^+})$ for $\delta$\,$>$\,$(\beta-\alpha)/2$~\cite{SM}. According to our definition, $\ket{\tilde \psi_A}$ is therefore a QSS. This analysis also gives
\begin{align}
    \big| \ket{\tilde \psi_A} - \ket{\psi_A}\big| = O(N^{\alpha/2-\beta/2+0^+}),\label{eq:statediff}
\end{align}
for $\ket{\psi_A}=\Pi_A\ket{\psi}/\sqrt{p_A}$, which implies $1-|\braket{\psi_A|\tilde\psi_A}|^2 = O(N^{\alpha-\beta+0^+})$. Meanwhile, the failure probability associated with the weakly nonunitary $\tilde R_{A,\phi}$ is upper bounded as $p_{\text{fail}} = O(N^{\alpha/2-\beta/2+0^+})$. 

Next we address the run-time, which is controlled by $d$ and $d' \ll [B \tau]^{-1}$. Since energy is extensive $||H||_{\infty} \sim N$ we require $\tau \sim N^{-1}$ and therefore the minimum polynomial degree required to resolve features on energy scale $B$ is $d' \sim N^{\beta+1+0^+}$. The query complexities relative to $\text{C}U_H$ and $U_{\psi}$, which we denote $q_{H}$ and $q_{\psi}$, respectively, then scale as 
\begin{align}
    q_{H} = \Theta(N^{5/4+\alpha/2+\beta+0^+}), \quad q_{\psi} = \Theta(N^{1/4+\alpha/2+0^+}). \label{eq:depth}
\end{align}
Recall also that $ q_{\psi}/2$ multiqubit Toffoli gates are required. The query complexities in Eq.~\eqref{eq:depth} should be compared with measurement-based schemes \cite{yang2020quantum,vasilyev2020monitoring}. If using QSP to engineer $Q'(e^{-iH \tau}) \approx \Pi_A$ with the same blur and polynomial degree $d'$ as above, QSSs would be created with probability $p_A$. In that case $q_H = \Theta(N^{3/2+\alpha+\beta})$ and $q_{\psi}=\Theta(N^{1/2+\alpha})$. Comparison with Eq.~\eqref{eq:depth} shows that QSVT provides a speedup for narrow energy windows.

\emph{Resolving dynamics} --- We now consider the extraction of the contributions $\braket{\psi_A|\mathcal{O}|\psi_{A'}}$ to the QSS decomposition. Crucially, as a consequence of Eq.~\eqref{eq:statediff}, we can achieve good approximations to $\braket{\psi_A|\mathcal{O}|\psi_{A'}}$ by probing the states $\ket{\tilde \psi_A}$. The route to estimating the diagonal terms $A = A'$ is then immediate: we prepare $\ket{\tilde \psi_A}$ and subsequently measure $\mathcal{O}$.

To measure the $A \neq A'$ contributions one introduces an additional ``register'' qubit, and uses this to construct blurred approximations to a register-controlled FP search $\ket{0}\bra{0}\otimes F_{\psi,A} + \ket{1}\bra{1}\otimes  F_{\psi,A'}$.
This can be readily accomplished since the choice of $A$ in our algorithm is determined solely by the rotation angles in QSP, which can now be coherently controlled by the register. 
Starting from a product state $\ket{0}\ket{\psi}$ of register and system, acting on the register with a Hadamard gate, and then evolving both register and system under the register-controlled FP search, we approximately prepare $(\ket{0} \ket{\psi_A}$\,$+$\,$\ket{1}\ket{\psi_{A'}})/\sqrt{2}$. Measuring $X$\,$\otimes$\,$\mathcal{O}$ and $Y$\,$\otimes$\,$\mathcal{O}$ (in different runs of the experiment) we can estimate the expectation value of $(X+iY)$\,$\otimes$\,$\mathcal{O}$, which is simply $\braket{\psi_A|\mathcal{O}|\psi_{A'}}$.

\begin{figure}
\includegraphics[width=0.485\textwidth]{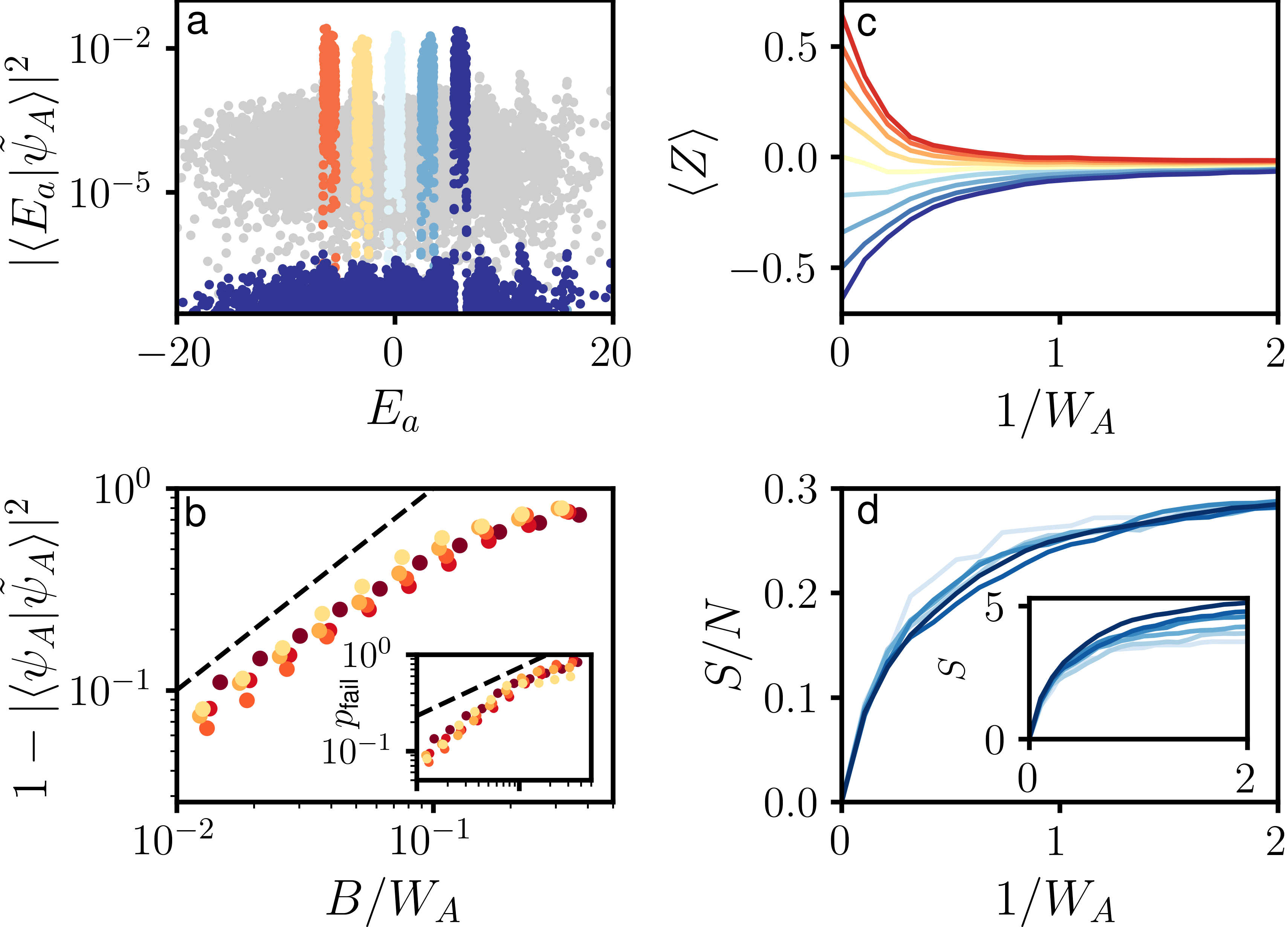}
\caption{
Numerical simulations. (a) Narrowly peaked eigenstate populations for five QSSs $\ket{\tilde \psi_A} \propto \tilde F_{\psi,A}\ket{\psi}$ with $E_A=-6,-3,0,3,6$ and $W_A=1$, derived from an initial  state $\ket{\psi}$ that has widely spread energy (grey). (b) Fidelity reduction (main) and failure probability (inset) for $W_A=1,2,\ldots,5$ (dark to light) as a function of $B$. The dashed black lines show the scalings (main) $\sim B/W_A$ and (inset) $\sim \sqrt{B/W_A}$ [see below Eq.~\eqref{eq:statediff}]. (c) Expectation values of a local observable $\langle Z\rangle$ for QSSs $\ket{\psi_A}\propto\Pi_A\ket{\psi}$ derived from various $\ket{\psi}$. In (a-c) the system size $N=18$. (d) Half-chain von Neumann entanglement entropy for $N=13,14,\ldots,18$ (light to dark) in $\ket{\psi_A}$; here $\ket{\psi}$ is the $N$-fold tensor product of a fixed single-qubit state. The main panel shows the entanglement entropy rescaled by $1/N$, and the inset shows the uncollapsed data. See~\cite{SM} for details.}
\label{fig:ising}
\end{figure}

\emph{Quantum Ising model} --- We now investigate the structure of QSSs in a paradigmatic model for chaotic many-body dynamics, the mixed-field Ising model 
with standard parameters 
from Ref.~\cite{banuls2011strong} and periodic boundary conditions.
We focus on the sector that is invariant under translation and reflection, and to access this sector we start from translation-invariant product states $\ket{\psi}$ with zero energy density $\braket{H}=0$. For our QSVT parameters we choose $\Delta^2$\,$=$\,$10^{-3}$  and $p_*$\,$=$\,$0.1$\,$\times$\,$W_A / \sqrt{\braket{H^2}}$~\cite{SM}.

Figure~\ref{fig:ising}a summarizes a proof-of-principle demonstration. We show the eigenstate overlaps of a single initial state, as well as of QSSs generated from it. We see clearly that the approximately unitary operator $\tilde F_{\psi,A}$ concentrates the amplitude within the target windows. In Fig.~\ref{fig:ising}b we compute the fidelity of the QSS $\ket{\tilde\psi_A} \propto \tilde F_{\psi,A}\ket{\psi}$ with $\ket{\psi_A} \propto \Pi_A \ket{\psi}$ (main), as well as the failure probability (inset), for various $W_A$ and $B$.
In the ideal FP search algorithm the former is guaranteed to be within $\Delta^2$ of unity, but this is not the case in our algorithm for $B$\,$\neq$\,$0$.
Our results show that for small blurring $B/W_A$ the reduction in fidelity is proportional to $p_B/p_A \sim B/W_A$, consistent with the bound in Eq.~\eqref{eq:statediff}.
The inset confirms our result that the failure probability has an upper bound $\sim \sqrt{p_B/p_A}$.

In Fig.~\ref{fig:ising}c,d we explore the properties of ideal QSSs, ignoring the above errors for simplicity.
Figure~\ref{fig:ising}c shows $\braket{\psi_A|Z|\psi_A}$, as a function of $A$ for various QSSs produced from several product states $\ket{\psi}$ having different initial $\braket{Z}$, and with fixed $\braket{H}=0$. As $A$ decreases, the expectation values all converge, well before $A$ reaches the Heisenberg energy scale. This behavior is consistent with the expectation that thermalization occurs within polynomial time, and with the bounds in Ref.~\cite{dymarsky2019new}. 

In Fig.~\ref{fig:ising}d we calculate the von Neumann entanglement entropy, $S$, of half chains of size $\lceil N/2\rceil$. We restrict to the initial state with $\braket{H}$\,$=$\,$0$ and $\braket{Z}$\,$\approx$\,$-0.5$ used in Fig.~\ref{fig:ising}c, and calculate $S$ for various $N$. Decreasing $W_A$ clearly increases the entanglement entropy \cite{banuls2020entanglement}, and at small $W_A$ the collapse of $S/N$ indicates volume-law entanglement, consistent with the anticipated exponential classical memory required to represent a thermalized pure state. However, the observed growth of $S$ on decreasing $W_A$ is rather slow compared with the upper bound $S \sim 1/W_A$ \cite{banuls2020entanglement}, corroborating a recent analysis of energy-filtered states \cite{morettini2024energy}.
This raises the question of precisely when the reconstruction of dynamics via the QSS decomposition is beyond the reach of classical computers.

\emph{Discussion} --- Theories for many-body quantum dynamics are often based on eigenstates; the ETH is a notable example, and forms the modern foundation of quantum statistical mechanics \cite{dalessio2016quantum,deutsch2018eigenstate}. However, there is a fundamental complexity barrier to creating eigenstates, and therefore to testing these theories \cite{atia2017fast}. Here we have shown how to efficiently and deterministically create QSSs having inverse polynomial energy width. These states can be used in alternative theoretical approaches based on the QSS decomposition. Furthermore, we expect that, by combining this decomposition with classical shadows \cite{huang2020predicting}, a quantum computer can generate an efficient classical representation of dynamics which is accurate up to polynomial times \cite{SM}.

Regarding near-term applications, our algorithm would be useful where conventional theory provides limited information, such as where predictions within canonical and microcanonical ensembles differ \cite{barre2001inequivalence,brandao2015equivalence,tasaki2018local,kuwahara2020gaussian}, e.g. at thermal phase transitions \cite{barre2001inequivalence,schuckert2023observation}, and where slow relaxation is observed, e.g. in systems exhibiting prethermalization \cite{mori2018thermalization} or weak ergodicity breaking \cite{moudgalya2022quantum}. Moreover, since our algorithm is deterministic, it can be used as a subroutine to build more advanced probes of many-body systems, for example to compute transport properties of materials such as spin and energy diffusion coefficients.

\emph{Acknowledgements}---The authors are grateful to Anurag Anshu, Mari-Carmen Ba\~{n}uls, Isaac Chuang, Anatoly Dymarsky, Vedika Khemani, Lin Lin, Quynh Nguyen, Thomas Schuster, and Peter Zoller for useful discussions. SJG was supported by QuantEmX grant GBMF9616 from ICAM, the Gordon and Betty Moore Foundation, and the U.S. Department of Energy, Office of Science, Office of High Energy Physics, under QuantISED Award DE-SC0019380. 
SC acknowledges partial support from the Center for Ultracold Atoms (an NSF Physics Frontiers Center; PHY-1734011), NSF CAREER (DMR-2237244), and the Heising-Simons Foundation (grant \#2024-4851).
This work was completed in part at the Aspen Center for Physics, which is supported by National Science Foundation grant PHY-2210452.

\newpage
\onecolumngrid
\newpage

\section*{Supplemental Material}
\setcounter{equation}{0}
\setcounter{figure}{0}
\setcounter{table}{0}
\setcounter{page}{1}
\setcounter{section}{0}
\renewcommand{\theequation}{S\arabic{equation}}
\renewcommand{\thefigure}{S\arabic{figure}}
\renewcommand{\thesection}{S\Roman{section}}
\renewcommand{\thepage}{S\arabic{page}}

This Supplemental Material is organized as follows. First, in Sec.~\ref{sec:oracle}, we describe in detail the polynomial approximations $\tilde R_{A,\phi}$ to the generalized reflections $R_{A,\phi}$ about energy windows $\mathcal{H}_A$. Using these results, in Sec.~\ref{sec:errors} we derive rigorous bounds on the errors which arise in our algorithm as a consequence of the fact that $\tilde R_{A,\phi} \neq R_{A,\phi}$. We then discuss the quasistationary state (QSS) decomposition in Sec.~\ref{sec:QSS}. In Sec.~\ref{sec:QSSerror} we bound the errors in this decomposition, and in Sec.~\ref{sec:QSSshadow} we show how classical shadows can be leveraged to analyze the dynamics of observables and entanglement. Finally in Sec.~\ref{sec:numerics} we provide additional details relevant to the numerical calculations in the main text, including the phases, or rotation angles, used for the fixed-point (FP) search algorithm. 

\section{Details of energy reflections}\label{sec:oracle}

Here we discuss the polynomial approximations $\tilde R_{A,\phi}$ to the generalized reflection operators $R_{A,\phi}$. The operators $\tilde R_{A,\phi}$ are diagonal in the energy basis by construction, and as we have discussed in the main text, it is convenient to express their entries as values of a smooth function $\tilde r_{A,\phi}(\theta)$, i.e. $\tilde r_{A,\phi}(E_a\tau) = \braket{E_a|\tilde \Pi_{A,\phi}|E_a}$. 
 We can express $\tilde{r}_{A,\phi}$ in terms of the polynomial coefficients as presented in the main text, which is reproduced here for convenience:
\begin{align}
	\tilde r_{A,\phi}(\theta) = \sum_{\ell=-d'}^{d'} q_{\phi,\ell} e^{-i\ell \theta}.
\end{align}
The barrier to efficiently constructing $\Pi_{A,\phi}$ is evident from this expression: for $d' = \text{poly}(N)$ it is not possible to design $\tilde{r}_{A,\phi}(\theta)$ that changes arbitrarily sharply, which is desired for an ideal $r_{A,\phi}(\theta)$. In other words, one cannot resolve features in the spectrum on energy scales that are exponentially small in $N$, and hence it is not possible to exactly resolve the sharp edges of the window $\mathcal{H}_A$. Moreover, if we simply fix $q_{\phi,\ell}$ to be the Fourier coefficients for the box function $r_{A,\phi}(\theta)$ presented in the main text, i.e. if we set
\begin{align}
    q_{\phi,\ell} = \frac{2i}{\pi \ell} \sin [\phi]\sin[\ell W_A \tau/2]e^{i\ell E_A \tau} + \delta_{0\ell} \Big[ e^{-i\phi} + \frac{i}{\pi}W_A\tau \sin \phi\Big]
\end{align}
and truncate the series at order $d'=\text{poly}(N)$, the resulting function $\tilde r_{A,\phi}(E)$ will suffer from the Gibbs phenomenon. That is, for any finite $d'$ the function $\tilde r_{A,\phi}(E)$ defined by the truncated Fourier series will feature a finite overshoot at the edges of the window $\mathcal{H}_A$. Since unitarity constrains us to have $|Q'(e^{-i\theta})|\leq 1$ \cite{motlagh2024generalized}, such an overshoot would warrant a rescaling of the polynomial, and a naive rescaling of this kind would cause $1-|Q'(e^{-i\theta})|$ to deviate significantly from unity over a finite range of $\theta$ for all finite $d'$. This would increase the typical failure probability to an $N$-independent constant at each of the $d$ steps of the FP search, and hence the success probability of the algorithm would decrease exponentially with $d$.

Instead, $\tilde r_{A,\phi}(\theta)$ must be a `smooth' approximation to $r_{A,\phi}(\theta)$. Convolving $r_{A,\phi}(\theta)$ with a Gaussian $g_B(\theta)=[2\pi B^2\tau^2]^{-1/2}e^{-\theta^2/(2B^2\tau^2)}$ having energy width $B \ll A$ we can suppress high order Fourier coefficients. We then define $\tilde r_{A,\phi}(A)$ from this truncated series (up to normalization). The function $\tilde r_{A,\phi}(A)$ does not suffer from the Gibbs phenomenon. It is also convenient to increase the width of the window by $\sim B$, and so we set
\begin{align}
	q_{\phi,\ell}/\eta = \frac{2i}{\pi \ell}\sin [\phi] \sin[\ell (W_A+hB) \tau/2] e^{-(\ell B \tau)^2/2 + i\ell E_A\tau} + \delta_{0\ell} \Big[ e^{-i\phi} + \frac{i}{\pi}(W_A+hB)\tau \sin \phi\Big], \label{eq:qphi}
\end{align}
for $\ell \neq 0$, while $q_{0,\phi}/\eta = e^{-i\phi} + i(W_A+hB)(\tau/\pi)\sin[\phi]$. The parameter $h$ should be chosen so that (i) $hB\ll W_A$ and (ii) $e^{-h^2/2}$ is much smaller than all amplitude errors; a simple choice is $e^{-h^2/2}\ll\Delta$. A possible rescaling $\eta$, which we introduce to ensure that $||\tilde R_{A,\phi}||_{\infty} \leq 1$, can be determined by bounding the truncation error in the Fourier series as follows,
\begin{align}
	|\sum_{|\ell| > d'}^{\infty} q_{\phi,\ell} e^{-i \theta \ell}| &\leq \frac{4}{\pi}\sum_{\ell=d'+1}^{\infty} \frac{1}{\ell} e^{-(\ell B \tau)^2/2}\notag\\
	& \leq \frac{4}{\pi (d'+1)} e^{-[(d'+1)B \tau]^2/2} \sum_{\ell' = 0}^{\infty} e^{-\ell'(d'+1)[B \tau]^2/2} \notag\\
	& \leq \frac{4}{\pi (d'+1)} \frac{e^{-[(d'+1)B \tau]^2/2}}{1-e^{(d'+1)[B \tau]^2/2}} =: \eta-1.
\end{align}
In the first line we used the triangle inequality, and in the second line we have bounded the sum of Gaussians by an infinite geometric series. This leads to the final line which specifies our choice of $\eta$.

\section{Errors in Quasi-stationary state preparation}\label{sec:errors}

Here we derive rigorous bounds on the errors which arise in our algorithm. To set up the analysis, it is useful to divide the Hilbert space as $\mathcal{H} = \mathcal{H}_{\bar{B}} \oplus \mathcal{H}_{B}$, where $\mathcal{H}_{B}$ is defined as the span of eigenstates with energies within $hB/2$ of either of $E_A \pm (W_A/2 + hB/2)$, i.e. the span of the set of $\ket{a}$ for which either of the conditions $|E_a-(E_{A} \pm [W_A+hB]/2)| \leq  hB/2$ holds. Note that, for $\ket{a} \in \mathcal{H}_{\bar{B}}$, we have $1-|\braket{a|\tilde R_{A,\phi}|a}| \leq e^{-h^2/8}$ and that, by definition, $\mathcal{H}_A \in \mathcal{H}_{\bar{B}}$. The value of $h$, which we choose to obtain favorable bounds on the errors, will be discussed below.

Let us define $\Pi_B$ and $\Pi_{\bar{B}}$ as the projectors into $\mathcal{H}_B$ and $\mathcal{H}_{\bar{B}}$, respectively. We also define the squared amplitude $p_{B} = |\Pi_{B} \ket{\psi}|^2$. For suitable $d$ and $d'$ we will show that the amplitude error $\sqrt{\epsilon}$ is upper bounded as
\begin{align}
	\sqrt{\epsilon}=|\Pi_{\bar{A}}\ket{\tilde \psi_A}| = O( (p_{B}/p_A)^{1/2} |\log \Delta|) \label{eq:finalerror}
\end{align}
in the regime where the right-hand side is much smaller than unity but nevertheless large compared with $\Delta$. For example, for $W_A = \Theta(N^{-\alpha})$, $B = O(N^{-\beta})$, and $\Delta=\Theta(N^{-\delta})$, this corresponds to $2\delta > \beta-\alpha > 0$, and our bound on the error is inverse polynomial in $N$. We will see that this bound arises from the (possibly coherent) addition of $d \sim |\log(\Delta)|/\sqrt{p_A}$ errors in amplitude, each of which is of order $\sqrt{p_{B}}$. To achieve the above scaling we will set $h=[8\mu \log(N)]^{1/2}$ with $\mu > \beta-\alpha/2+1/4$.

Recall that the ideal FP search operator takes the form \cite{yoder2014fixed,gilyen2019quantum,martyn2021grand}
\begin{align}
    F = R_{\psi,\phi_{d-2}}R_{A,\phi_{d-3}}\cdots R_{\psi,\phi_1}R_{A,\phi_0}, \label{eq:appF}
\end{align}
where, for brevity, on the left-hand side we have omitted subscripts $A$ and $\psi$ labelling the energy window $\mathcal{H}_A$ and initial state $\ket{\psi}$, respectively, i.e. $F \equiv F_{\psi,A}$. The set of phases $\vec{\phi}=(\phi_0,\phi_1,\ldots)$ is chosen such that $|F\ket{\psi}-\ket{\psi_A}| \leq \Delta$.

The many-body level spacing at the edge of the energy window $\mathcal{H}_A$ is in general exponentially small in $N$, and therefore we cannot efficiently resolve the edges of the window $A$. Our approximation to $\tilde F \equiv \tilde F_{\psi,A}$ to $F$ arises from replacing $R_{A,\phi_k}$ with $\tilde R_{A,\phi_k}$. The operator $R_{A,\phi_k}$ acts as $e^{i\phi_k}$ within $\mathcal{H}_A$ and as $e^{-i\phi_k}$ in its complement, while $\tilde R_{A,\phi_k}$ interpolates between these two behaviors over an energy interval $B \ll W_A$. Explicitly, 
\begin{align}
    \tilde F = R_{\psi,\phi_{d-2}}\tilde R_{A,\phi_{d-3}}\cdots R_{\psi,\phi_1}\tilde R_{A,\phi_0}. \label{eq:appFtilde}
\end{align}
To keep track of the errors it is convenient to write the initial state as
\begin{align}
	\ket{\psi} &= (1-p_{B})^{1/2}\ket{\psi_{\bar{B}}} + p_{B} \ket{\psi_{B}},
\end{align}
where $\sqrt{p_B} \ket{\psi_B} = \Pi_B \ket{\psi}$, i.e. $\ket{\psi_B} \in \mathcal{H}_B$ and $\ket{\psi_{\bar{B}}} \in \mathcal{H}_{\bar{B}}$ are normalized states. We also define generalized reflection operators about $\ket{\psi_{\bar{B}}}$, the component of $\ket{\psi}$ supported away from the edges of the window,
\begin{align}
	R_{\psi_{\bar{B}},\phi} = e^{i\phi(2\Pi_{\psi_{\bar{B}}}-\mathbbm{1})},
\end{align}
where $\Pi_{\psi_{\bar{B}}} = \ket{\psi_{\bar{B}}}\bra{\psi_{\bar{B}}}$. With these objects in hand, we define $F_{\bar{B}}$ and $\tilde F_{\bar{B}}$ through the replacement of all operators $R_{\psi,\phi}$ with $R_{\psi_{\bar{B}},\phi}$ in Eqs.~\eqref{eq:appF} and \eqref{eq:appFtilde}, respectively, i.e.
\begin{align}
	 F_{\bar{B}} &= R_{\psi_{\bar{B}},\phi_{d-2}}R_{A,\phi_{d-3}}\cdots R_{\psi_{\bar{B}},\phi_1}R_{A,\phi_0},\label{eq:appFB}\\
	 \tilde F_{\bar{B}} &= R_{\psi_{\bar{B}},\phi_{d-2}}\tilde R_{A,\phi_{d-3}}\cdots R_{\psi_{\bar{B}},\phi_1}\tilde R_{A,\phi_0}. \notag
\end{align}
We now outline our overall strategy and categorize the various errors which arise. First, we can trivially rewrite
\begin{align}
    \tilde F \ket{\psi} = F_{\bar{B}}\ket{\psi_{\bar{B}}} + (\tilde F \ket{\psi} -F_{\bar{B}}\ket{\psi_{\bar{B}}}). \label{eq:appsplit1}
\end{align}
Since $F_{\bar{B}}$ implements the ideal FP search transforming $\ket{\psi_{\bar{B}}}$ into $\mathcal{H}_A$ we can apply existing error bounds \cite{yoder2014fixed} to characterize this contribution to $\tilde F \ket{\psi}$. We will then require rigorous bounds on the term in parentheses in Eq.~\eqref{eq:appsplit1}. This term is rewritten
\begin{align}
	\tilde F\ket{\psi} - F_{\bar{B}}\ket{\psi_{\bar{B}}}
	= \tilde F\big(\ket{\psi} - \ket{\psi_{\bar{B}}}\big) + \big(\tilde F - \tilde F_{\bar{B}}\big)\ket{\psi_{\bar{B}}} + \big(\tilde F_{\bar{B}} - F_{\bar{B}}\big)\ket{\psi_{\bar{B}}}\label{eq:appsplit2} 
\end{align}
The magnitude of the left-hand side is then upper bounded, via the triangle inequality, by the sum of the magnitudes of the three terms on the right-hand side of Eq.~\eqref{eq:appsplit2}. Combining the resulting bounds with the errors incurred when using $F_{\bar{B}}$ to transform $\ket{\psi_{\bar{B}}}$ into $\mathcal{H}_A$, we will arrive at our final error bounds. First we discuss the bound on $|\Pi_{\bar{A}}F_{\bar{B}}\ket{\psi_{\bar{B}}}|$, which follows from Ref.~\cite{yoder2014fixed}. We then derive bounds on the three rows in Eq.~\eqref{eq:appsplit2} in the order displayed.

The error $|\Pi_{\bar{A}}F_{\bar{B}}\ket{\psi_{\bar{B}}}|$ is fixed by $\Delta$, an input to the search algorithm which determines the number of steps $d$. From the definition we have
\begin{align} |\Pi_{\bar{A}}F_{\bar{B}}\ket{\psi_{\bar{B}}}| \leq \Delta, \label{eq:appbound0}
\end{align}
so for $\Delta = O(N^{-\delta})$ we have an inverse polynomial bound. The number of steps $d$ is proportional to $\log(1/\Delta)$ for small $\Delta$, so $d \sim \delta \log N$. We can therefore choose a large constant $\delta$ at only multiplicative cost in runtime.

Next, the magnitude of $\tilde F(\ket{\psi} - \ket{\psi_{\bar{B}}})$ is bounded by $|\ket{\psi}-\ket{\psi_{\bar{B}}}|$ since $||\tilde F||_{\infty} \leq 1$. From the definition of $\ket{\psi_{\bar{B}}}$ we have $|\ket{\psi}-\ket{\psi_{\bar{B}}}| = O(p_{B}^{1/2})$ for $p_{B} \ll 1$ which gives
\begin{align}
    |\tilde F(\ket{\psi} - \ket{\psi_{\bar{B}}})| = O(p_{B}^{1/2}), \label{eq:appbound1}
\end{align}
as our bound on the first term on the right-hand side of Eq.~\eqref{eq:appsplit2}.

The magnitude of $\tilde F\ket{\psi_{\bar{B}}} - \tilde F_{\bar{B}}\ket{\psi_{\bar{B}}}$ is bounded by $||\tilde F - \tilde F_{\bar{B}}||_{\infty}$. This quantifies the difference arising from the replacement of $R_{\psi,\phi}$ with $R_{\psi_{\bar{B}},\phi}$. First write
\begin{align}
    R_{\psi,\phi} = R_{\psi_{\bar{B}},\phi} + 2p_{B}^{1/2}(\cos\phi) \epsilon_{\psi}, \label{eq:appepspsi}
\end{align}
where
\begin{align}
    \epsilon_{\psi} = p^{1/2}_B\Big(\ket{\psi_B}\bra{\psi_B}-\ket{\psi_{\bar{B}}}\bra{\psi_{\bar{B}}}\Big) +(1-p_{B})^{1/2}\Big(\ket{\psi_B}\bra{\psi_{\bar{B}}}+\text{H.c.}\Big)
\end{align}
has eigenvalues $\pm 1$. Inserting Eq.~\eqref{eq:appepspsi} into the expression for $\tilde F_{\bar{B}}$ in Eq.~\eqref{eq:appFB} gives 
\begin{align}
    \tilde F-\tilde F_{\bar{B}} = \prod_{k=0}^{(d-3)/2} \Pi_{\psi,\phi_{2k+1}}\tilde\Pi_{A,\phi_{2k}} - \prod_{k=0}^{(d-3)/2} \big(\Pi_{\psi,\phi_{2k+1}}+2p_{B}^{1/2}(\cos\phi_{2k+1})\epsilon_{\psi}\big)\tilde\Pi_{A,\phi_{2k}},
\end{align}
and we then expand the resulting expression in powers of $p_{B}^{1/2}$,
\begin{align}
     \tilde F-\tilde F_B = \sum_{k=1}^{(d-3)/2} p_{B}^{k/2} \mathcal{A}_k,
\end{align}
which defines the operators $\mathcal{A}_k$ implicitly. Each term $\mathcal{A}_k$ is a sum of ${(d-3)/2 \choose k}$ products of operators, with each product involving $k$ factors of $\epsilon_{\psi}\tilde \Pi_{A,\phi}$ and $(d-3)/2-k$ factors of $\Pi_{\psi,\phi'} \tilde \Pi_{A,\phi}$, as well as products of cosines. We can therefore use the triangle inequality and Holder's inequality to write
\begin{align}
	||\mathcal{A}_k||_{\infty} \leq &{(d-3)/2 \choose k} ||\epsilon_{\psi,\phi}||^k_{\infty}  ||\Pi_{\psi,\phi}||^{(d-3)/2-k}_{\infty} ||\tilde \Pi_{A,\phi}||^{(d-3)/2}_{\infty}. 
\end{align}
Using $||\epsilon_{\psi}||_{\infty} = 1$, $||\Pi_{\psi,\phi}||_{\infty} \leq 1$ and $||\tilde \Pi_{A,\phi}||_{\infty} \leq 1$ (which must hold for any $\tilde \Pi_{A,\phi}$ embedded in a block of a unitary), this becomes $||\mathcal{A}_k||_{\infty} \leq {(d-3)/2 \choose k}$. The error is therefore
\begin{align}
	||\tilde F-\tilde F_{\bar{B}}||_{\infty} \leq \big[1+ p_{B}^{1/2}\big]^{(d-3)/2} -1,
\end{align}
so for $d p_{B}^{1/2} \ll 1$ we have
\begin{align}
    |(\tilde F - \tilde F_{\bar{B}})\ket{\psi_{\bar{B}}}| = O(d p_{B}^{1/2}) \label{eq:appbound2}
\end{align}
as our bound on the magnitude of the second term in the right-hand side of Eq.~\eqref{eq:appsplit2}.

Finally, the difference $(\tilde F_{\bar{B}}- F_{\bar{B}})\ket{\psi_{\bar{B}}}$ quantifies the effect of replacing $R_{A,\phi}$ with $\tilde R_{A,\phi}$ in the case where the initial state $\ket{\psi_{\bar{B}}}$ has support only away from the edges of the window. The analysis of this contribution is simplified by the fact that $\tilde R_{A,\phi}$ is diagonal in the basis of energy eigenstates. Therefore
\begin{align}
    |(\tilde F_{\bar{B}}- F_{\bar{B}})\ket{\psi_{\bar{B}}}| \leq ||\Pi_{\bar{B}}(\tilde F_{\bar{B}}- F_{\bar{B}})\Pi_{\bar{B}}||_{\infty}.
\end{align}
Next, note that 
\begin{align}
    \Pi_{\bar{B}}\tilde F_{\bar{B}}\Pi_{\bar{B}} = \prod_{k=0}^{(d-3)/2} R_{\psi_{\bar{B}},\phi_{2k+1}}\Pi_{\bar{B}}\tilde R_{A,\phi_{2k}}\Pi_{\bar{B}},
\end{align}
and similar for $\Pi_{\bar{B}}F_{\bar{B}}\Pi_{\bar{B}}$. That is, if we are only interested in the actions of $F_{\bar{B}}$ and $\tilde F_{\bar{B}}$ projected into $\mathcal{H}_{\bar{B}}$, we can replace the generalized energy reflections (and our approximations to them) by their projections into $\mathcal{H}_{\bar{B}}$. This simplification follows from $\Pi_B^2 = \Pi_B$, $[R_{\psi_{\bar{B}},\phi},\Pi_B]=0$, $[R_{A,\phi},\Pi_B]=0$, and $[\tilde R_{A,\phi},\Pi_B]=0$.
The error $(\tilde F_{\bar{B}}- F_{\bar{B}})\ket{\psi_{\bar{B}}}$ is therefore controlled only by the differences between $\Pi_{\bar{B}}\tilde R_{A,\phi_{2k}}\Pi_{\bar{B}}$ and $\Pi_{\bar{B}} R_{A,\phi_{2k}}\Pi_{\bar{B}}$. Let us write
\begin{align}
    \Pi_{\bar{B}}\tilde R_{A,\phi}\Pi_{\bar{B}} = \Pi_{\bar{B}} R_{A,\phi}\Pi_{\bar{B}} + \epsilon_{\bar{B},\phi} \label{eq:appepsbarB}
\end{align}
which defines the error $\epsilon_{\bar{B},\phi}$. For an approximation $\tilde R_{A,\phi}$ to $R_{A,\phi}$ obtained by Gaussian smoothing, we have $||\epsilon_{\bar{B},\phi}||_{\infty}=e^{-h^2/8}$. Inserting Eq.~\eqref{eq:appepsbarB} into the expression for $\tilde F_{\bar{B}}$ in Eq.~\eqref{eq:appFB} we find
\begin{align}
    \Pi_{\bar{B}} \tilde F_{\bar{B}}\Pi_{\bar{B}} &= \Pi_{\bar{B}}\prod_{k=0}^{(d-3)/2} R_{\psi_{\bar{B}},\phi_{2k+1}}\big[R_{A,\phi_{2k}}+\epsilon_{\bar{B},\phi_{2k}}\big]\Pi_{\bar{B}}\notag\\
    &= \sum_{k=0}^{(d-3)/2} e^{-h^2 k/8} \mathcal{B}_k,
\end{align}
where we have defined $\mathcal{B}_k$ as the operators appearing in the expansion in powers of $e^{-h^2/8}$. We then have
\begin{align}
    ||\Pi_{\bar{B}}(\tilde F_{\bar{B}}- F_{\bar{B}})\Pi_{\bar{B}}||_{\infty} \leq \sum_{k=0}^{(d-3)/2}e^{-h^2 k/8}||\mathcal{B}_k||_{\infty}.
\end{align}
From Holder's inequality and the triangle inequality we have $||\mathcal{B}_k||_{\infty} \leq {(d-3)/2 \choose k}$ since $\mathcal{B}_k$ is a sum of ${(d-3)/2 \choose k}$ products of operators, each of which has norm upper bounded by unity. This gives
\begin{align}
    ||\Pi_{\bar{B}}(\tilde F_{\bar{B}}- F_{\bar{B}})\Pi_{\bar{B}}||_{\infty} \leq [1+e^{-h^2/8}]^{(d-3)/2}-1,
\end{align}
and from this we find
\begin{align}
    |(\tilde F_{\bar{B}}- F_{\bar{B}})\ket{\psi_{\bar{B}}}| = O(d e^{-h^2/8}) \label{eq:appbound3}
\end{align}
for small $de^{-h^2/8}$. This is our bound on the magnitude of the third and final term on the right-hand side of Eq.~\eqref{eq:appsplit2}.

Combining the above results we find a bound on amplitude of $\tilde F\ket{\psi}$ outside of the target energy window $\mathcal{H}_A$. Introducing numerical constants $c_j$ with $j=1,2,3$ we write the right-hand side of Eq.~\eqref{eq:appbound1} as $c_1p_{B}^{1/2}$, of Eq.~\eqref{eq:appbound2} as $c_2 d p_{B}^{1/2}$, and of Eq.~\eqref{eq:appbound3} as $c_3 d e^{-h^2/8}$,
\begin{align}
    |\tilde F\ket{\psi} - F_{\bar{B}}\ket{\psi_{\bar{B}}}| \leq (c_1+c_2 d) p_{B}^{1/2} + c_3 d e^{-h^2/8}. \label{eq:apppsidiff}
\end{align}
We also have the error displayed in Eq.~\eqref{eq:appbound0}. Acting with $\Pi_{\bar{A}}$ on both sides of Eq.~\eqref{eq:appsplit2} we then find
\begin{align}
    |\Pi_{\bar{A}} \tilde F_{\psi,A}\ket{\psi}| \leq \Delta +  (c_1+c_2 d) p_{B}^{1/2} + c_3 d e^{-h^2/8} \label{eq:appsumerr}
\end{align}
for $d p_{B}^{1/2} \ll 1$ and $d e^{-h^2/8} \ll 1$.

We now discuss the $N$ dependence of the various contributions to Eq.~\eqref{eq:appsumerr}. Given a tolerance $\Delta = O(N^{-\delta})$ and a lower bound $p_*$ on $p_A$, the number of steps $d$ is determined by $d \geq 2\lceil \ln(2/\Delta)/(2\sqrt{p_*})\rceil+1$. For $W_A = \Theta(N^{-\alpha})$ and initial energy width $W_{\psi}=\Theta(N^{-1/2})$ we assume $p_A =\Theta(N^{-1/2-\alpha})$ and therefore set $p_* = \Theta(N^{-1/2-\alpha})$. This scaling results in $d = \Theta( N^{\alpha/2+1/4}\log N)$ at large $N$. Recall that $d$ is proportional to $\delta$, meaning that increasing $\delta$ is far less expensive than decreasing the width of the window. 

Consider next the $c_2$ term in Eq.~\eqref{eq:appsumerr}. This quantity is small for $p_{B} \ll d^{-2}$, and we assume $p_{B} \sim B/W_{\psi}$, or more explicitly $p_B = O(N^{-1/2-\beta+0^+})$, where we have introduced a small positive constant $0^+$ to absorb logarithmic factors. We can therefore set $B = O(N^{-\beta})$ with $\beta > \alpha$. The total energy width of the subspace $\mathcal{H}_B$ is $2hB$, and we demand that (i) this width is asymptotically smaller than $W_A$, and also that (ii) the $c_3$ term in Eq.~\eqref{eq:appsumerr} is small relative to the other contributions. That is, we require $hB \ll W_A$ and $d e^{-h^2/8} \ll 1$. The first of these conditions gives an upper bound $h = O(N^{\beta-\alpha})$ and, using $d = \Theta( N^{\alpha/2+1/4+0^+})$, the second condition is a lower bound $h = \Omega(\log^{1/2}(N))$. Clearly these conditions are consistent, so let us choose the energy width $hB$ of  $\mathcal{H}_B$ by setting $h = [8\mu \log(N)]^{1/2}$ for a constant $\mu > \beta-\alpha/2+1/4$. This choice of $h$ ensures that the $c_3$ term in Eq.~\eqref{eq:appsumerr} is negligible compared with the $c_2$ term. 

In summary, with window width $W_A = \Theta(N^{-\alpha})$, blur $B = O(N^{-\beta})$ with $\beta > \alpha$, and tolerance $\Delta = \Theta(N^{-\delta})$ with $2\delta > \beta-\alpha$, we can ensure that the amplitude error is upper bounded as
\begin{align}
    |\Pi_{\bar{A}} \tilde F_{\psi,A}\ket{\psi}| = O(N^{\alpha/2-\beta/2+0^+})
\end{align}
at large $N$, where $0^+$ is a positive constant that captures logarithmic factors. 

These results allow us to straightforwardly bound the failure probability, which is given by $1-|\tilde F\ket{\psi}|^2$. Above we have bounded $|\tilde F \ket{\psi}-F_{\bar{B}}\ket{\psi_{\bar{B}}}|$, and $\Pi_B F_{\bar{B}}\Pi_B$ is unitary by construction, so $|F_{\bar{B}}\ket{\psi_{\bar{B}}}|=1$. Using $|\tilde F \ket{\psi}-F_{\bar{B}}\ket{\psi_{\bar{B}}}|^2 \geq |\tilde F \ket{\psi}|^2 - 2|\tilde F \ket{\psi}|+1$ and our bound on the left-hand side of this equality, which follows from the sums of Eqs.~\eqref{eq:appbound1}, \eqref{eq:appbound2}, and \eqref{eq:appbound3}, we find that the failure probability is
\begin{align}
    1-|\tilde F \ket{\psi}|^2 = O(d p_{B}^{1/2})
\end{align}
for $d p_{B}^{1/2} \ll 1$ and in the regime discussed above (where the $c_2$ term dominates the right-hand side of Eq.~\eqref{eq:appsumerr}). Inserting the $N$-dependence of $p_A$ and $p_{B}$ we then have
\begin{align}
     1-|\tilde F_{\psi,A} \ket{\psi}|^2 = O(N^{(\alpha-\beta)/2+0^+}).
\end{align}
It is clear that this bound is not very restrictive, and we expect that it can be improved. For example, the fact that $|\braket{E_a|\tilde R_{A,\phi}|E_a}|$ is well below unity for $E_a$ near the edges of the window implies that, if the first implementation of a generalized reflection $R_{A,\phi}$ succeeds, the support of the state in $\mathcal{H}_B$ is reduced, and this suppresses the probability of later failures. Moreover, in the main text we have shown numerically that the failure probability scales as $\sim B/A$ rather than $\sim \sqrt{B/A}$.

Finally, we address the difference between the QSS $\ket{\tilde \psi_A}=\tilde F \ket{\psi}/|\tilde F \ket{\psi}|$ that we create and the ideal QSS $\ket{\psi_A} = \Pi_A \ket{\psi}/\sqrt{p_A}$. First observe that
\begin{align}
	| \ket{\tilde \psi_A} - \ket{\psi_A} | = | \tilde F \ket{\psi} - \ket{\psi} | + O(dp_B^{1/2}),
\end{align}
where we have inserted the definition of $\ket{\tilde \psi_A}$ and used a binomial expansion $|\tilde F \ket{\psi}|^{-1} = 1+O(dp_B^{1/2})$. We then bound the first term on the right-hand side using the triangle inequality
\begin{align}
    | \tilde F \ket{\psi} - \ket{\psi_A}| \leq | \tilde F \ket{\psi} - F_{\bar{B}}\ket{\psi_{\bar{B}}}| + | F_{\bar{B}}\ket{\psi_{\bar{B}}} - \ket{\psi_A}|. \label{eq:stateerr}
\end{align}
The first term on the right-hand side has been bounded above in Eq.~\eqref{eq:apppsidiff}, with the dominant contribution to the bound equal to $c_2 d p_B^{1/2}$. Since $\mathcal{H}_A \in \mathcal{H}_{\bar{B}}$ by definition, the second term is upper bounded by $\Delta$, and this contribution is negligible compared with $dp_B^{1/2}$ in the regime of interest. We can then conclude that
\begin{align}
    | \ket{\tilde \psi_A} - \ket{\psi_A}| = O(N^{\alpha/2-\beta/2+0^+}), \quad |\braket{\psi_A|\tilde \psi_A}|^2 = O(N^{\alpha-\beta+0^+}),
\end{align}
where the bound on the fidelity in the second expression follows from the first expression. This result implies that expectation values in $\ket{\tilde \psi_A}$ and in $\ket{\psi_A}$ are close. More generally, for a bounded observable $\mathcal{O}$ we have
\begin{align}
   \big| \braket{\tilde \psi_A|\mathcal{O}|\tilde \psi_{A'}}-\braket{\psi_A|\mathcal{O}|\psi_{A'}}\big| = O(N^{\alpha/2-\beta/2+0^+}),
\end{align}
which also holds for $A=A'$. Therefore, the QSSs generated by our algorithm can be used as the basis for a `QSS decomposition' of the dynamics, as discussed in the main text. We elaborate on this idea in the next section.

\section{Quasi-stationary state decomposition}\label{sec:QSS}
The decomposition of an initial state $\ket{\psi}$ into quasistationary states (QSSs) $\ket{\psi_A} = \Pi_A\ket{\psi}/\sqrt{p_A}$ supported within different energy windows of width $W=W_A$ allows us to analyze dynamics out to times $t \sim W^{-1}$. In particular, if we divide the many-body spectrum into nonoverlapping windows $A$ having widths $W = W_A$ and centers $E_{A+1}=E_A+W$, we can define an approximation to the Hamiltonian of the form $H_W = \sum_{A} E_A \Pi_A$. The dynamics generated by $H_W$ corresponds to neglecting energy differences between eigenstates within the window, e.g. for a pure initial state it corresponds to the approximation $\ket{\psi} \approx \sum_A \sqrt{p_A} \ket{\psi_A} e^{-iE_A t}$. This is the (pure state) QSS decomposition of dynamics introduced in the main text. In Sec.~\ref{sec:QSSerror} we provide bounds on the errors which arise when dynamics is reconstructed from the QSS decomposition, and in Sec.~\ref{sec:QSSshadow} we show how classical shadows \cite{huang2020predicting} of the different contributions can be extracted. 

First let us provide an overview of the idea. Suppose that we are interested in analyzing the dynamics of a density matrix of a region $R$. Starting from an initial density matrix $\rho$ of the entire system, the time-evolved density matrix $\rho(t)=e^{i H t}\rho e^{-i H t}$. Expectation values of observables supported in $R$, as well as entanglement properties of this subregion, are encoded in $\text{Tr}_{\bar{R}}\rho(t)$, where $\bar{R}$ is the complement of $R$. We define the (pure or mixed state) QSS decomposition of the dynamics as the set of objects $\text{Tr}_{\bar{R}}[\Pi_A \rho \Pi_{A'}]$ for all pairs of windows $\mathcal{H}_A$ and $\mathcal{H}_{A'}$ for which $\text{Tr}_{\bar{R}}[\Pi_A \rho \Pi_{A'}]$ is non-negligible. Since $||H||_{\infty} = O(N)$ and $W$ is inverse polynomial in $N$, there are polynomially many pairs of windows. Therefore, for $R$ with polynomial Hilbert space dimension, e.g. for a subregion of $O(\log N)$ qubits, the full set of objects $\text{Tr}_{\bar{R}}[\Pi_A \rho \Pi_{A'}]$ occupies classical memory that is at most polynomial in $N$.

Given the QSS decomposition, the dynamics of the quantum system, as well as the relation between dynamics and spectral properties, can therefore be studied in classical post-processing. We can view the QSS decomposition for a small subregion $R$ as a memory-efficient representation of spectral properties which is available even when $N$ is too large for $H$ to be diagonalized exactly. Crucially, we can use it to approximate $\text{Tr}_{\bar{R}}\rho(t) \approx \text{Tr}_{\bar{R}}\rho_W(t)$, where
\begin{align}
    \rho_W(t) = e^{-iH_W t}\rho e^{iH_W t} = \sum_{AA'}\text{Tr}_{\bar{R}}\big[ \Pi_A \rho \Pi_{A'}] e^{-i(E_A-E_{A'})t}.
\end{align}
From $\rho_W(t)$ we can then estimate observable expectation values as well as entanglement measures such as the purity of $R$,
\begin{align}
    \text{Tr}_{\bar{R}}\rho_{W}(t) = \sum_{AA'}  \text{Tr}_{\bar{R}}\big[\Pi_A \rho \Pi_{A'}\big] e^{-i(E_A-E_{A'})t}. \label{eq:QSSrho}
\end{align}
This expression can be straightforwardly generalized to entanglement measures involving higher powers of the density matrix as well as to different trace structures.

\subsection{Errors in the decomposition}\label{sec:QSSerror}

For an density matrix $\rho$ (pure or mixed) the replacement of $H$ with $H_W$ leads us to an approximation $\rho_W(t)=e^{-i H_W t} \rho e^{i H_W t}$ to $\rho(t) = e^{-i H t}\rho e^{i H t}$. Defining $g_W = W^{-1}(H-H_W)$, we can quantify the error by the trace distance
\begin{align}
    ||\rho(t)-\rho_W(t)||_1 &= ||e^{-iWt g_W}\rho_W(t)e^{i W t g_W} -\rho_W(t)||_1  \notag\\
    &= \sum_{m,n \atop m \neq 0 \lor n \neq 0} \frac{(-1)^m(i W t)^{m+n}}{m!n!} g_W^m \rho_W(t) g_W^n.
\end{align}
Using the triangle inequality and Holder's inequality we then have
\begin{align}
     ||\rho(t)-\rho_W(t)||_1 \leq \sum_{m,n \atop m \neq 0 \lor n \neq 0} \frac{(Wt)^{m+n}}{m!n!}||\rho_W(t)||_1 ||g_W||_{\infty}^{m+n}.
\end{align}
Since the eigenvalues of $g_W$ are in $[-1/2,1/2]$, and $||\rho_W(t)||_1 = 1$, the error is 
\begin{align}
     ||\rho(t)-\rho_W(t)||_1 \leq e^{Wt}-1 = O(Wt),
\end{align}
where the asymptotic notation on the right-hand side corresponds to the limit $Wt \to 0$.

Based on the data processing inequality, it can be shown that a bound on the trace distance between many-body density matrices implies a bound on the error in estimates for local observables. In particular,  
\begin{align}
     \frac{1}{2}||\rho(t)-\rho_W(t)||_1 = \text{sup}_{\Pi} \text{Tr}\big[ (\rho(t)-\rho_W(t)) \Pi\big],
\end{align}
where $\Pi$ is a projection operator. The above bound shows that, as expected, the QSS with energy resolution $W$ allows for an accurate reconstruction of dynamics for times $t \ll W^{-1}$, where the error is bounded by $Wt$.

\subsection{Classical shadows of the decomposition}\label{sec:QSSshadow}
Having established the validity of a QSS decomposition let us now discuss its extraction in more detail. For simplicity we here consider pure $\rho=\ket{\psi}\bra{\psi}$. To estimate the diagonal terms in the QSS decomposition, i.e. $\text{Tr}_{\bar{R}}\big[\Pi_A \rho \Pi_{A}\big]$, one can use our QSS preparation algorithm to generate an approximation to $\ket{\psi_A}$. Following this, classical shadows can be extracted for the region $R$. In practice this corresponds to applying random basis rotations and then measuring qubits in $R$ in the computational basis. Denoting the shadows by $\sigma_r$, where the index $r$ labels runs of the experiment, the average over these shadows is
\begin{align}
    \mathbbm{E}_r\big[ \sigma_r \big] = \text{Tr}_{\bar{R}}\ket{\psi_A}\bra{\psi_A} = \frac{\text{Tr}_{\bar{R}}[\Pi_A \rho\Pi_A]}{p_A},
\end{align}
where we neglect the difference between $\ket{\psi_A}$ and $\ket{\tilde \psi_A}$ for simplicity. Extracting $\text{Tr}_{\bar{R}}[\Pi_A \rho\Pi_A]$ itself requires that we also measure $p_A$. This can be achieved using quantum phase estimation or amplitude estimation. Off-diagonal terms $\text{Tr}_{\bar{R}}\big[\Pi_A \rho \Pi_{A'}\big]$ with $A \neq A'$ can be determined by synthesizing classical shadows with an idea discussed in the main text. Introducing a single `register' qubit, from the initial state
\begin{align}
    \frac{1}{\sqrt{2}}\big(\ket{0}+\ket{1}\big) \ket{\psi}
\end{align}
of register and system, the register-controlled FP search $\Pi_{0} F_{\psi,A} + \Pi_{1} F_{\psi,A'}$ generates
\begin{align}
    \frac{1}{\sqrt{2}}\big(\ket{0}\ket{\psi_A}+\ket{1}\ket{\psi_{A'}}\big)
\end{align}
up to inverse polynomial error. Now in each run $r$ of the experiment one measures either Pauli $X$ or Pauli $Y$ on the register (with equal probability) and extracts a classical shadow from region $R$ of the system. If measuring $X$ we set a boolean variable $a_r=0$ and if we measure $Y$ we set $a_r=1$. The result of this ($X$ or $Y$) measurement is denoted $b_r = \pm 1$. The matrix-valued shadow of the subregion $R$ that we extract in this run is denoted $\sigma_r$. It can then be verified that averaging $2 i^{a_r} b_r \sigma_r$ over runs of such experiments yields 
\begin{align}
    \mathbbm{E}_r \big[ 2 i^{a_r} b_r \sigma_r] = \text{Tr}_{\bar{R}} \ket{\psi_A}\bra{\psi_{A'}} = \frac{\text{Tr}_{\bar{R}}\big[ \Pi_A \rho \Pi_{A'}]}{\sqrt{p_A p_{A'}}}.
\end{align}
This concludes our discussion of the QSS decomposition. Although we have here discussed the extraction of the QSS decomposition using our algorithm based on QSVT, measurement-based schemes are also viable. 

\section{Details of numerical simulations}\label{sec:numerics}

Here we provide additional information relevant to the numerical demonstration of the algorithm presented in the main text. The initial states $\ket{\psi}$ are pure translation-invariant product states, and we restrict ourselves to $\braket{\psi|H|\psi}=0$ with respect to the Ising Hamiltonian $H = -\sum_j (Z_j Z_{j+1}+hZ_j+gX_j)$ with $g=-1.05$, $h=0.5$ \cite{banuls2011strong}, and $N=18$ qubits. Since we are working with product states we have $\braket{X}^2+\braket{Y}^2+\braket{Z}^2=1$, where expectation values are with respect to the initial $\ket{\psi}$. The condition $\braket{H}=0$ then implies a relation between $\braket{X}$ and $\braket{Z}$, i.e. $\braket{Z}^2+h\braket{Z}+g\braket{X}=0$. Restricting to initial states with $\braket{Y}\geq 0$, a choice of $\braket{Z}$ therefore specifies both $\braket{X}$ and $\braket{Y}$.

In Figure 2a in the main text, we start from a single initial state $\ket{\psi}$ with $\braket{Z}=0$ and construct $\tilde F_{\psi,A}\ket{\psi}/|\tilde F_{\psi,A}|$ for five different windows $E_A=-6,-3,0,3,6$ having widths $W_A=1$. There we set $\Delta^2=10^{-3}$, and $p_*=0.1 \times W_A/\sqrt{\braket{H^2}}$, where since $\braket{H}=0$ the energy variance $\braket{H^2}=N(g^2+h^2)$. These parameters determine the minimum degree for QSVT $d_*=2\lceil \ln(2/\Delta)/(2\sqrt{p_*})\rceil+1$, and we use this minimum degree in practice, i.e. $d=d_*$. Given $p_*$ and $d$ the phases $\phi_j$ necessary for the FP search algorithm are given in Ref.~\cite{yoder2014fixed,li2024revisiting} as
\begin{align}
    \phi_k = 2(-1)^k \cot^{-1}\big( \sqrt{p_*} \tan[(k+1)\pi/d] \big).\label{eq:QSVTphases}
\end{align}
This is sufficient to define the FP algorithm $F_{\psi,A}$ with idealized energy reflections $R_{A,\phi_j}$, where $F_{\psi,A}\ket{\psi} \approx \ket{\psi_A}$ up to a (computable) global phase. However, given polynomial time we can only implement an approximation $\tilde R_{A,\phi_j}$ to $R_{A,\phi}$.
To specify $\tilde R_{A,\phi_j}$, we must choose $\tau$ such that its product with the many-body bandwidth (in the sector invariant under translation and reflection) is less than $2\pi$. Based on a conservative estimate for the bandwidth, we fix $\tau=\pi/[2N(1+g+h)]$. We must additionally specify the parameters $h$ and $B$. Our error analysis has revealed that for $d \gg 1$ the two dominant amplitude errors scale as $\sim d e^{-h^2/8}$ and $d \sqrt{p_B} \propto \sqrt{p_B/p_A}$. Since $d < 10^2$ in our simulations, we set $h=8$ to suppress the first kind of error. To suppress the latter we set $B = b/[d_*^2 \tau]$ with $b=1$ in Fig.~2(a), while we vary $b$ in Fig. 2(b). Our approximations $\tilde R_{A,\phi}$ to $R_{A,\phi}$ are then constructed using a $(2d'+1)$-term Laurent polynomial in $e^{-iH\tau}$, with $d'=5/[B\tau]$. 

\end{document}